\documentclass[english,prb,notitlepage,twocolumn]{revtex4-1}
\usepackage[T1]{fontenc}
\usepackage[latin9]{inputenc}
\usepackage{babel}
\usepackage{float}
\usepackage{amsmath}
\usepackage{graphicx}
\usepackage{xcolor}
\usepackage[unicode=true,pdfusetitle,
 bookmarks=true,bookmarksnumbered=false,bookmarksopen=false,
 breaklinks=false,pdfborder={0 0 1},backref=false,colorlinks=false]
 {hyperref}


\begin{document}
%\title{Convergence behavior of ghost rotationally-invariant slave-boson theory at small bath size: a comparative study with dynamical mean-field theory}
%\title{Convergence of  quantum embedding approaches with small bath sizes: ghost-rotationally-invariant slave-boson and dynamical mean-field theory}
\title{Accuracy of ghost-rotationally-invariant slave-boson and dynamical mean field theory as a function of the impurity-model bath size}
%\title{Accuracy of quantum-embedding approaches as a function of the impurity-model bath size: ghost-rotationally-invariant slave-boson and dynamical mean-field theory}

\author{Tsung-Han Lee$^{1}$, Nicola Lanat\`a$^{2}$, Gabriel Kotliar$^{1,3}$}
\affiliation{$^{1}$Physics and Astronomy Department, Rutgers University, Piscataway,
New Jersey 08854, USA}
\affiliation{$^{2}$School of Physics and Astronomy, Rochester Institute of Technology,
84 Lomb Memorial Drive, Rochester, New York 14623, USA}
\affiliation{$^{3}$Condensed Matter Physics and Materials Science Department,
Brookhaven National Laboratory, Upton, New York 11973, USA}
\begin{abstract}
We compare the accuracy of the ghost-rotationally-invariant slave-boson (g-RISB) theory and dynamical mean-field theory (DMFT) on the single-band Hubbard model, as a function of the number of bath sites in the embedding impurity Hamiltonian. Our benchmark calculations confirm that the accuracy of g-RISB can be systematically improved by increasing the number of bath sites, similar to DMFT. With a few bath sites, we observe that g-RISB is systematically more accurate than DMFT for the ground-state observables.
On the other hand, the relative accuracy of these methods is generally comparable for the quasiparticle weight and the spectral function. %(both at low and high energies). 
As expected, we observe that g-RISB satisfies the variational principle in infinite dimensions, as the total energy decreases monotonically towards the exact value as a function of the number of bath sites, suggesting that the g-RISB wavefunction may approach the exact ground state in infinite dimensions. 
Our results suggest that the g-RISB is a promising method for first principle simulations of strongly correlated matter, which can capture the behavior of both static and dynamical observables, at a relatively low computational cost.
%ghost-orbitals in the bath of the embedded impurity model, similar to DMFT. Moreover, we demonstrate that g-RISB generally produces more accurate total energy and double occupancy than DMFT at small bath sizes, while DMFT gives more accurate quasiparticle weight. Both methods produce accurate physical observables with five or more bath orbitals.
%, where the total energy and double occupancy in g-RISB converge rapidly with bath size Nb = 3 while DMFT requires Nb = 5 to reach the same level of convergence. 
%In addition, g-RISB captures reliable spectral functions comparable to DMFT at low and high frequencies. Our results demonstrate that g-RISB is a promising method that requires a few bath orbitals to capture accurate static and dynamic observables.
\end{abstract}
\maketitle

\section{Introduction}

Quantum embedding approaches have recently attracted significant attention in %multidisciplinary fields of 
condensed matter physics, material science, and quantum chemistry~\cite{DMFT_RMP_1996,DMFT_RMP_2006,Maier_QCT,Knizia2012,Wouters,QE_Sun,Lecherman_2007,Lanata_2015_PRX,Potthoff_VCA,SEET,Zgid_2017}. The common idea of these approaches consists in mapping the original interacting lattice to an embedded quantum impurity model, whose parameters are determined self-consistently by matching the properties of the impurity model and the lattice. The dynamical mean-field theory (DMFT) is the first example of these approaches~\cite{DMFT_RMP_1996}, which has become a standard method for strongly correlated materials~\cite{DMFT_RMP_2006}. Nevertheless, the calculations of the dynamical Green's function in DMFT can be time-consuming for realistic multiorbital systems. Therefore, significant effort has been put into developing more efficient quantum embedding techniques~\cite{Knizia2012,Wouters,Bulik_DET,Lanata_2015_PRX,Lanata_2017_PRL,RISB_DMET_2018,SOET}.

The rotational-invariant slave-boson (RISB) mean-field theory and other related approaches are among the most efficient methodologies for studying strongly correlated systems \citep{Lecherman_2007,Bunemann_2007,Lanata_2008,Medici2005,Yu2012,Georgescu2015,Lanata_2015_PRX,Lanata_2017_PRL,RISB_DMET_2018,Knizia2012}. Unlike DMFT, these frameworks are often classified as static quantum embedding approaches, as the embedded impurity model parameters are determined self-consistently by matching the lattice and the impurity density matrix, instead of the local Green's function and self energy. These static quantum embedding approaches are particularly successful in capturing the static observables and low-energy spectral functions in strongly correlated systems in qualitative agreement with more sophisticated DMFT\citep{Lecherman_2007,Lanata_2015_PRX,Lanata_2017_PRL}.

%The rotational-invariant slave-boson (RISB) mean-field theory and other related approaches are efficient methodologies for studying strongly correlated systems \citep{Lecherman_2007,Bunemann_2007,Medici2005,Yu2012,Georgescu2015,Lanata_2015_PRX,Lanata_2017_PRL,RISB_DMET_2018,Knizia2012}.
%These frameworks are often classified as static quantum embedding approaches that the original interacting models are mapped to a non-interacting reference system and an embedded quantum impurity model, whose parameters are determined self-consistently by matching the density matrix of the two reference systems. These static quantum embedding approaches are particularly successful in capturing the static observables and low energy spectral functions in strongly correlated systems in qualitative agreement with more sophisticated dynamical mean-field theory (DMFT)\citep{Lecherman_2007,Lanata_2015_PRX,Lanata_2017_PRL}.

Recently, the ghost-RISB (g-RISB) extension has been introduced, where
auxiliary ghost degrees of freedoms are added to the non-interacting
lattice model and the bath of the impurity model\citep{gRISB_2017}.
It has been shown that g-RISB with additional ghost orbitals can capture reliably both the spectral and the static observables of the infinite dimensional Hubbard and Anderson lattice models\citep{gRISB_2017,gRISB_2021,gRISB_2022,Guerci2019,Guerci_thesis}. 
At the same time,
a similar approach has been developed in the density matrix embedding theory ~\citep{Fertitta2018,Sriluckshmy2021,Si1994}, and the ancilla qubits technique~\citep{ancilla_quibits_2020}.
%and that DMET has a close connection to the RISB framework~\citep{Lanata_2015_PRX,RISB_DMET_2018,RISB_DMET_Lee_2019}.
%The Ew-DMET also introduces auxiliary degrees-of-freedom into the bath of the impurity model to fit the moments of the single-particle Green's function. 
%The accuracy of Ew-DMET can also be systematically improved by adding more auxiliary orbitals, which controls the number of poles of the Green's function or the self-energy ~\citep{Fertitta2018,Sriluckshmy2021,Si1994}.

The idea of increasing the bath size to improve the accuracy in g-RISB %and Ew-DMET 
reminisces the bath discretization of dynamical mean-field theory with exact-diagonalization impurity
solver (DMFT-ED)\citep{DMFT_RMP_1996,Caffarel_Krauth_1994,Rozenberg1994}. In DMFT-ED,
the continuous bath of the Anderson impurity model is fitted by a
small number of bath orbitals such that the impurity model can be
solved by ED.  %Moreover, it has been shown that the accuracy of DMFT-ED can be systematically improved by increasing the number of orbitals in the bath, and that five bath orbitals are enough for obtaining accurate calculations of the phase diagram of the  single-orbital Hubbard model compared to the numerical renormalization group simulations \citep{Tong2001}. The convergence behavior with increasing bath orbitals has also been demonstrated in multiorbital systems~\cite{Liebsch_2011}. 
In the past decade, DMFT-ED has been extensively applied to the cluster extensions of DMFT and multiorbital realistic materials~\cite{Liebsch_2005,Liebsch_2007,Liebsch_2010,Liebsch_2011,Civelli_2005,Kyung_2006,Capone2006,Capone2007,Weber2010}. Highly efficient exact-diagonalization techniques have also been developed~\cite{Zgid2012,Lin2013,Lu2014,Go2017,Mejuto-Zaera2019,Zhu2019}.
%While the convergence of DMFT-ED as a function of bath size has been studied~\cite{Tong2001,Liebsch_2005,Liebsch_2011}, the accuracy and the convergence behavior of g-RISB observables as a function of the bath size have not been systematically investigated yet. 
%Therefore, it is important to asses which method is more accurate and efficient. %Indeed,  this is particularly important for frontier problems in strongly correlated materials or chemical systems, which generally involve multiple orbitals and/or large clusters~\cite{Liebsch_2005,Liebsch_2007,Liebsch_2010,Liebsch_2011,Civelli_2005,Kyung_2006,Capone2006,Capone2007,Weber2010}. 

Since the Hilbert space of the embedded impurity model grows exponentially with the number of the bath degrees of freedom in all the quantum embedding approaches, it is important to assess which method is more accurate and efficient at smaller bath sizes.
Indeed,  this is particularly important for frontier problems in strongly correlated materials or chemical systems, which generally involve multiple orbitals and/or large clusters~\cite{Liebsch_2005,Liebsch_2007,Liebsch_2010,Liebsch_2011,Civelli_2005,Kyung_2006,Capone2006,Capone2007,Weber2010}. 
%In this respect, an appealing feature of the g-RISB method is that it only requires to compute the ground-state density matrix. Furthermore, it does not require a bath-fitting procedure, or using analytical continuation methods for extracting the spectral properties on the real axis~\cite{JARRELL1996133}. Therefore, by construction, g-RISB is numerically more efficient than DMFT-ED, for calculations performed using the same bath size.
Nevertheless, the accuracy and the convergence behavior of g-RISB observables as a function of the bath size have not been systematically investigated yet.
%%As g-RISB only requires the density matrix for self-consistency calculations and does not require a bath-fitting procedure, it is more efficient than DMFT-ED by construction. Nevertheless, the accuracy and the convergence behavior of g-RISB observables with increasing bath size are still unexplored.
%%As DMFT-ED, g-RISB, and Ew-DMET all display systematic improvement of accuracy
%with increasing number of bath orbitals in the impurity model \citep{Liebsch_2011,Fertitta2018,Sriluckshmy2021},
%%it is important to assess which method is more accurate and efficient at smaller bath size. It is an important question for determining suitable method to study frontier problems in strongly correlated materials or chemical systems, where the numbers of the physical and bath orbitals is constrained by the exponential growing Hilbert space.

In this work, we study the convergence behavior of g-RISB as a function of the number of bath orbitals in the single-band Hubbard model and compare it to DMFT-ED. Our results indicate that g-RISB generally provides us with more accurate energy and ground-state properties at small bath sizes, %while DMFT-ED yields more accurate quasiparticle weights. 
while the relative accuracy of these methods is comparable for the spectral properties.
%, and depends on the Hamiltonian's parameters. 
%In particular, for the ground-state observables, we show that g-RISB provides results with accuracy comparable with the ``exact'' value from the continuous-time quantum Monte Carlo solver (CTQMC)~\cite{CTQMC_RMP_2011} already for $N_b = 3$. On the other hand, within DMFT-ED, it is necessary to use  $N_{b}=5$ to reach comparable accuracy. Both methods require five or more bath orbitals to obtain accurate quasiparticle weights.
%In addition, we show that the g-RISB spectral function is in good agreement with the DMFT spectral function with ED and CTQMC solvers, both at low and high frequencies. 
Moreover, we verify numerically that g-RISB satisfies the variational principle in the limit of infinite coordination number, \emph{i.e.}, that the energy decreases monotonically towards the exact value, suggesting that the g-RISB wavefunction may approach the exact ground state in infinite dimensions.%. These results support the hypothesis that the g-RISB wavefunction approaches the exact ground state wavefunction in infinite dimensions.

%Our results demonstrate that g-RISB is a promising tool for %capturing reliable physical quantities of Hubbard models using a small size of bath orbitals. This promising feature opens a route to investigate the challenging effects of multiorbital and short-range correlations in realistic materials, combining the power of
%computing reliably the static and the dynamical properties of the of Hubbard models (in the limit of infinite coordination number), using a relatively small number of bath orbitals.
%This feature of the g-RISB makes it a promising framework for investigating strongly correlated materials. 
%Furthermore, the fact that this method only requires to compute ground state observables of the embedding Hamiltonian opens the possibility of employing highly efficient wavefunction-based techniques ~\cite{Zgid2012,Lin2013,Lu2014,Go2017,Mejuto-Zaera2019,Zhu2019,White1992,White1999,Chan2011,Cao_2021,Bauernfeind2017,Zhang1997,Zheng2017,Hao2019,hidden_fermion,Booth_AFQMC_EDMET}.

\section{Model and methods}

We consider a single-orbital Hubbard model on the Bethe lattice in
the limit of infinite coordination number \citep{DMFT_RMP_1996}:

\begin{align}
H & =\sum_{\mathbf{k}}\sum_{\sigma}\epsilon_{\mathbf{k}\sigma}c_{\mathbf{k}\sigma}^{\dagger}c_{\mathbf{k}\sigma}+\sum_{i}Un_{i,\uparrow}n_{i\downarrow},\label{eq:H}
\end{align}
where $c_{\mathbf{k}\sigma}^{\dagger}$ and $c_{\mathbf{k}\sigma}$
are the electron creation and annihilation operators for momentum
$\mathbf{k}$ and spin $\sigma$, $n_{i\sigma}=c_{i\sigma}^{\dagger}c_{i\sigma}$
is the number operator on site $i$, and $U$ is the Coulomb interaction.
The energy unit is set to $D=1$, where $D$ is the half bandwidth
of the semicircular density of states on the Bethe lattice.

\subsection{Ghost-rotationally-invariant slave-boson theory}

The g-RISB approach is utilized to study the static observables and the dynamical spectral function of the single-orbital Hubbard model. The detailed derivations of g-RISB are shown in Refs. \onlinecite{gRISB_2017,gRISB_2021,gRISB_2022}. Here, we briefly review the g-RISB formalism. 

The g-RISB formalism is entirely encoded in the following Lagrangian
\citep{gRISB_2022}:
%\begin{widetext}
\begin{align}
\mathcal{L}[\Phi, & E^{c};R,\lambda;D,\lambda^{c};\Delta,\Psi_{0},E]=\frac{1}{N}\langle\Psi_{0}|\hat{H}^{\text{qp}}[R,\lambda]|\Psi_{0}\rangle\nonumber\\ 
 &+E(1-\langle\Psi_{0}|\Psi_{0}\rangle) +\sum_{i}\Big[\langle\Phi_{i}|\hat{H}_{i}^{\text{emb}}[D,\lambda^{c}]|\Phi_{i}\rangle \nonumber\\
 &+E_{i}^{c}(1-\langle\Phi_{i}|\Phi_{i}\rangle)\Big] -\sum_{i}\Big[\sum_{ab}\big[\lambda_{i}+\lambda_{i}^{c}\big]_{ab}\big[\Delta_{i}\big]_{ab}\nonumber\\
 &+\sum_{ca\alpha}\big(\big[D_{i}\big]_{a\alpha}\big[R_{i}\big]_{c\alpha}\big[\Delta_{i}(\mathbf{1}-\Delta_{i})\big]_{ca}^{\frac{1}{2}}+\text{c.c.}\big)\Big],
\end{align}
%\end{widetext}
where $H_{\text{qp}}$ is the quasiparticle Hamiltonian
and $H_{\text{emb}}$ is the embedding Hamiltonian, and $|\Psi_0\rangle$ and $|\Phi_i\rangle$ are their wavefunction, respectively. The quasiparticle
Hamiltonian is as follows:
\begin{equation}
H^{\text{qp}}=\sum_{\mathbf{k}}\sum_{ab}\Big[\sum_{\sigma}R_{a\sigma}\epsilon_{\mathbf{k},\sigma}R_{\sigma b}^{\dagger}+\lambda_{ab}\Big]f_{\mathbf{k}a}^{\dagger}f_{\mathbf{k}b},
\end{equation}
where $\sigma\in\{\uparrow,\downarrow\}$ is the physical spin degrees
of freedom and $a,b$ corresponds to the auxiliary quasiparticle degrees
of freedom $f_{a}$, whose size can be systematically increased to
improve the accuracy of g-RISB. The matrices $\big[R_i\big]_{a\sigma}\equiv\sum_{b}\langle\Phi_i|c_{i\sigma}^\dagger c_{ib}| \Phi_i \rangle \big[\Delta_i(1-\Delta)_i\big]_{ba}^{-1/2}$ and $\lambda$ correspond
to the quasiparticle renormalization matrix and the renormalized potential,
respectively. The $\Delta_i$ corresponds to the local quasiparticle density matrix. In this work, we use up to seven auxiliary quasiparticle
spin-orbitals, i.e., $a,b\in\{1\uparrow,1\downarrow,....,7\uparrow,7\downarrow\}$.
Note that the minimal single quasiparticle-orbital, $a,b\in\{1\uparrow,1\downarrow\}$,
recovers the original RISB approach.

\begin{figure}[t]
\begin{centering}
\includegraphics[scale=0.32]{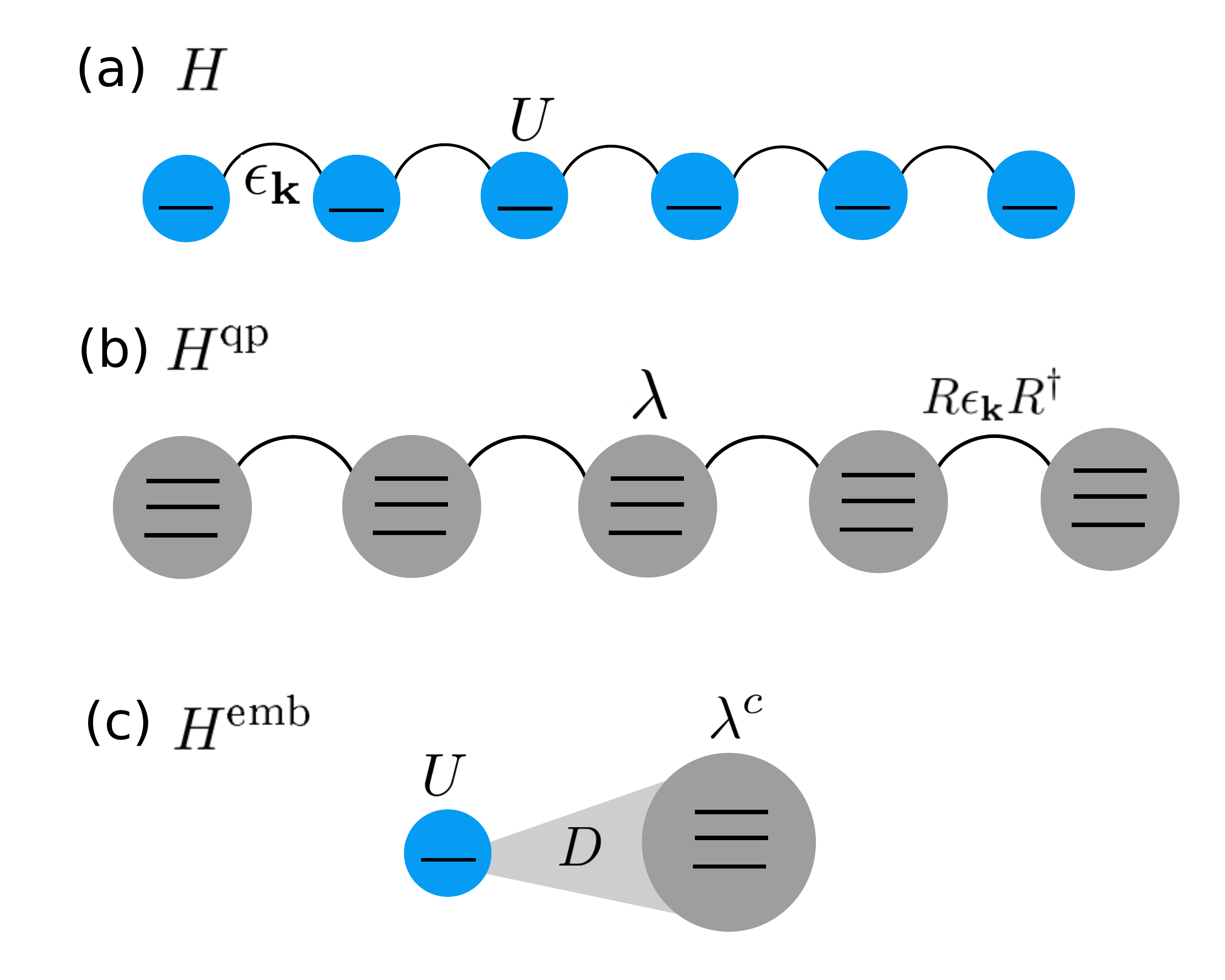}
\par\end{centering}
\caption{Schematic representation of the (a) original Hubbard model $H$, (b)
non-interacting quasiparticle Hamiltonian $H^{\text{qp}}$, and (c)
interacting embedding impurity model $H^{\text{emb}}$. We use three
bath sites, including the ghost orbitals, to illustrate the structure. \label{fig:scheme}}
\end{figure}

The embedding Hamiltonian has the following form:

\begin{align}
\hat{H_{i}}^{\text{emb}} & =U\hat{n}_{i\uparrow}\hat{n}_{i\downarrow}-\sum_{\sigma}\mu\hat{n}_{i\sigma}\nonumber \\
 & +\sum_{a\sigma}\Big(\big[D_{i}\big]_{a\sigma}\hat{c}_{i\sigma}^{\dagger}\hat{f}_{ia}+\text{h.c.}\Big)+\sum_{ab}\big[\lambda_{i}^{c}\big]_{ab}\hat{f}_{ib}\hat{f}_{ia}^{\dagger},
\end{align}
where $D$ and $\lambda^{c}$ describe the hybridization and the
bath potential, respectively. The schematic representation of the
two Hamiltonian is shown in Fig. \ref{fig:scheme}. Note the g-RISB
$H_{\text{emb}}$ is similar to the DMFT impurity model with the discretized
bath orbitals.

The two Hamiltonians are coupled with each other through the following
self-consistent g-RISB equations:
\begin{equation}
\frac{1}{\mathcal{N}}\Big[\sum_{\mathbf{k}}n_{f}(R\epsilon_{\mathbf{k}}R^{\dagger}+\lambda)\Big]_{ba}=\big[\Delta_{i}\big]_{ab}
\end{equation}
\begin{equation}
\frac{1}{\mathcal{N}}\Big[\sum_{\mathbf{k}}\epsilon_{\mathbf{k}}R^{\dagger}n_{f}(R\epsilon_{\mathbf{k}}R^{\dagger}+\lambda)\Big]_{\sigma a}=\sum_{ac\sigma}\big[D\big]_{c\sigma}\big[\Delta_{i}(\mathbf{1}-\Delta_{i})\big]_{ac}^{\frac{1}{2}}
\end{equation}
\begin{equation}
\sum_{cd\sigma}\frac{\partial}{\partial\big[\Delta_{i}\big]_{ab}}\Big(\big[\Delta(\mathbf{1}-\Delta)\big]_{cd}^{\frac{1}{2}}D_{d\sigma}R_{c\sigma}+\text{c.c.}\Big)+\big[\lambda+\lambda^{c}\big]_{ab}=0
\end{equation}
\begin{equation}
H_{i}^{\text{emb}}|\Phi_{i}\rangle=E^{c}|\Phi_{i}\rangle
\end{equation}
\begin{equation}
\langle\Phi_{i}|c_{i,\sigma}^{\dagger}f_{i,a}|\Phi\rangle-\sum_{c}\big[\Delta(\mathbf{1}-\Delta_{i})\big]_{ac}^{\frac{1}{2}}\big[R_{i}\big]_{c\sigma}=0
\end{equation}
\begin{equation}
\langle\Phi_{i}|f_{ib}f_{ia}^{\dagger}|\Phi_{i}\rangle-\big[\Delta_{i}\big]{}_{ab}=0
\end{equation}
where $n_f$ is the Fermi function, and the variables $R$, $\lambda$, $D$, $\lambda^{c}$ are determined
self-consistently. With the converged $R$ and $\lambda$, one can
compute the Green's function from 
\begin{equation}
G_{\sigma}(\mathbf{k},\omega)=R_{\sigma a}^{\dagger}\big[\omega+i0^{+}-R\epsilon_{\mathbf{k}}R^{\dagger}+\lambda\big]_{ab}^{-1}R_{b\sigma}
\end{equation}
and the self-energy can be determined from the Dyson equation
\begin{equation}
\Sigma_{\sigma}(\omega)=[G^{0}_{\sigma}(\mathbf{k},\omega)]^{-1}-[G_{\sigma}(\mathbf{k},\omega)]^{-1}.
\end{equation}
Note that the self-energy is momentum independent in g-RISB~\cite{gRISB_2017,gRISB_2021}. The quasiparticle
renormalization weight is determined from 
\begin{equation}
Z_{\sigma}=\big[1-\frac{\partial\text{Re}\Sigma_{\sigma}(\omega)}{\partial\omega}\big|_{\omega\rightarrow0}\big]^{-1}.
\end{equation}
In this work, we will focus on the paramagnetic solution so the spin
index $\sigma$ will be suppressed.

\begin{figure}[t]
\begin{centering}
\includegraphics[scale=0.4]{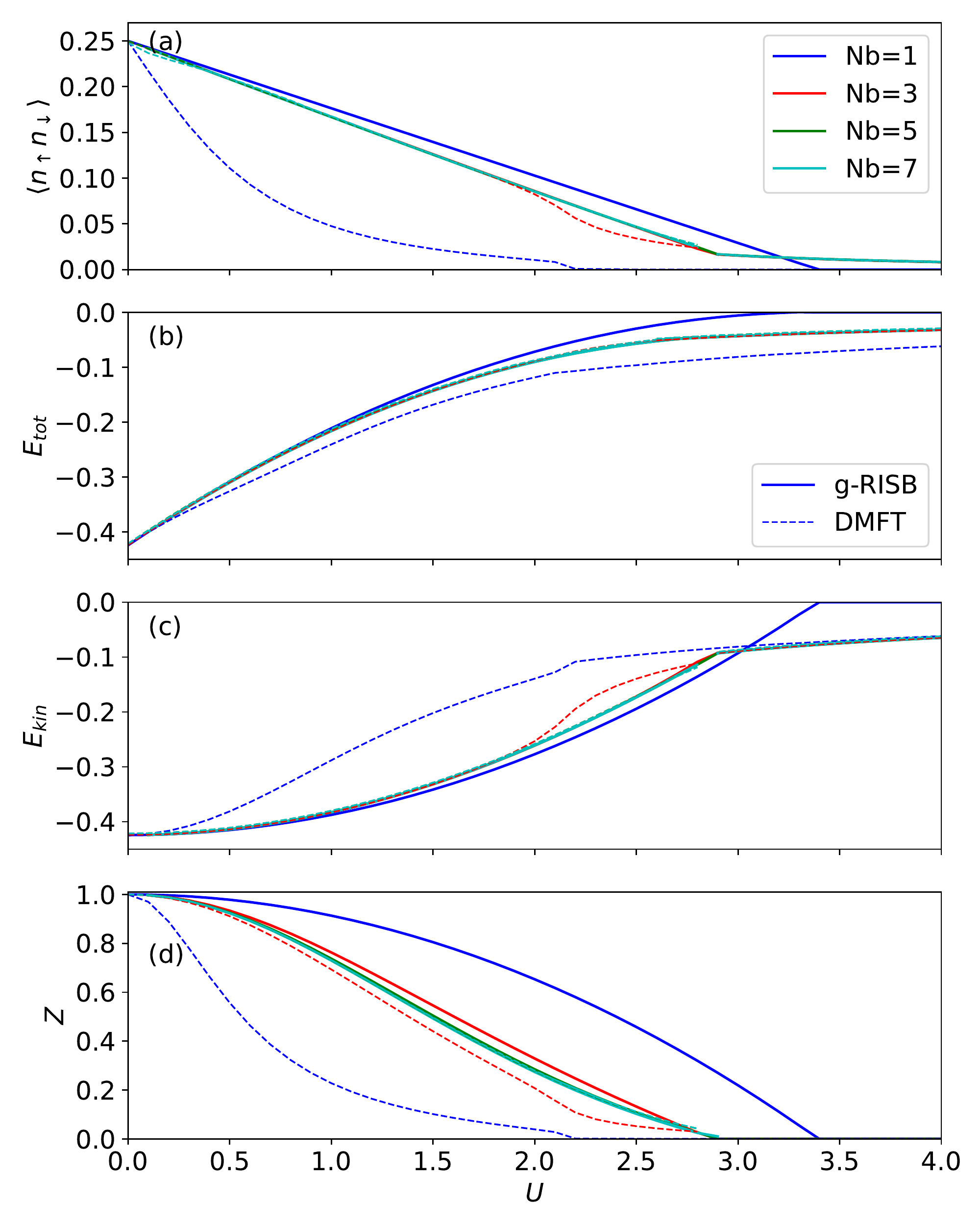}
\par\end{centering}
\caption{(a) Double occupancy $\langle n_{\uparrow}n_{\downarrow}\rangle$, (b) total energy $E_{\text{tot}}$, (c) kinetic energy $E_{\text{kin}}$, and (d) quasiparticle weight $Z$ as
a function of coulomb interaction $U$ with increasing bath size $N_{b}$
at half-filling. The g-RISB and DMFT are shown as solid lines and dashed
lines, respectively. \label{fig:Z_docc_E}}
\end{figure}

\subsection{Dynamical mean-field theory}

We apply the DMFT-ED algorithm with the discretized bath orbitals to address
the convergence of the bath size on the single-orbital Hubbard model%.
%The algorithm can be found in a large volume of literature
~\citep{Caffarel_Krauth_1994,Tong2001,Capone2007,Liebsch_2011,Strand_2011}. In particular, we use the Lanczos algorithm to solve for the ground state wavefunction and Green's function~\citep{Caffarel_Krauth_1994,Liebsch_2007}.
For the bath discretization algorithm, we introduce a fictitious inverse temperature $\beta=200$ to fit the hybridization function on the Matsubara frequency. For $N_{b}=1$,
we use the weight function $1/i\omega_{n}$ to obtain better occupancy
and fitting~\cite{Capone2007}. For $N_{b}>1$, we use the uniform weight function. The
number of frequency points used for the hybridization function fitting
is $N_{\omega_{\text{max}}}=200$, and the conjugate-gradient method
is utilized for the minimization. In our calculations, we found that
the hybridization function fitting is essential to obtain reasonable
total energy for $N_{b}\leq 3$. This feature is not reported in the literature,
where most studies focus on the spectral function, quasiparticle
weight, and the double occupancy \citep{Caffarel_Krauth_1994,Tong2001,Capone2007,Liebsch_2011,Strand_2011}. %We also performed DMFT calculations with CTQMC solver~\cite{CTQMC_RMP_2011} and maximum entropy analytic continuation~\cite{JARRELL1996133} to obtain numerically-exact solutions.

%For the completeness of our benchmark, 
We also performed DMFT calculations with CTQMC solver~\cite{CTQMC_RMP_2011}, which gives us numerically-exact solutions on the Bethe lattice. The maximum entropy method is utilized for the analytic continuation of  Green's functions~\cite{JARRELL1996133}.

\begin{figure}[t]
\begin{centering}
\includegraphics[scale=0.4]{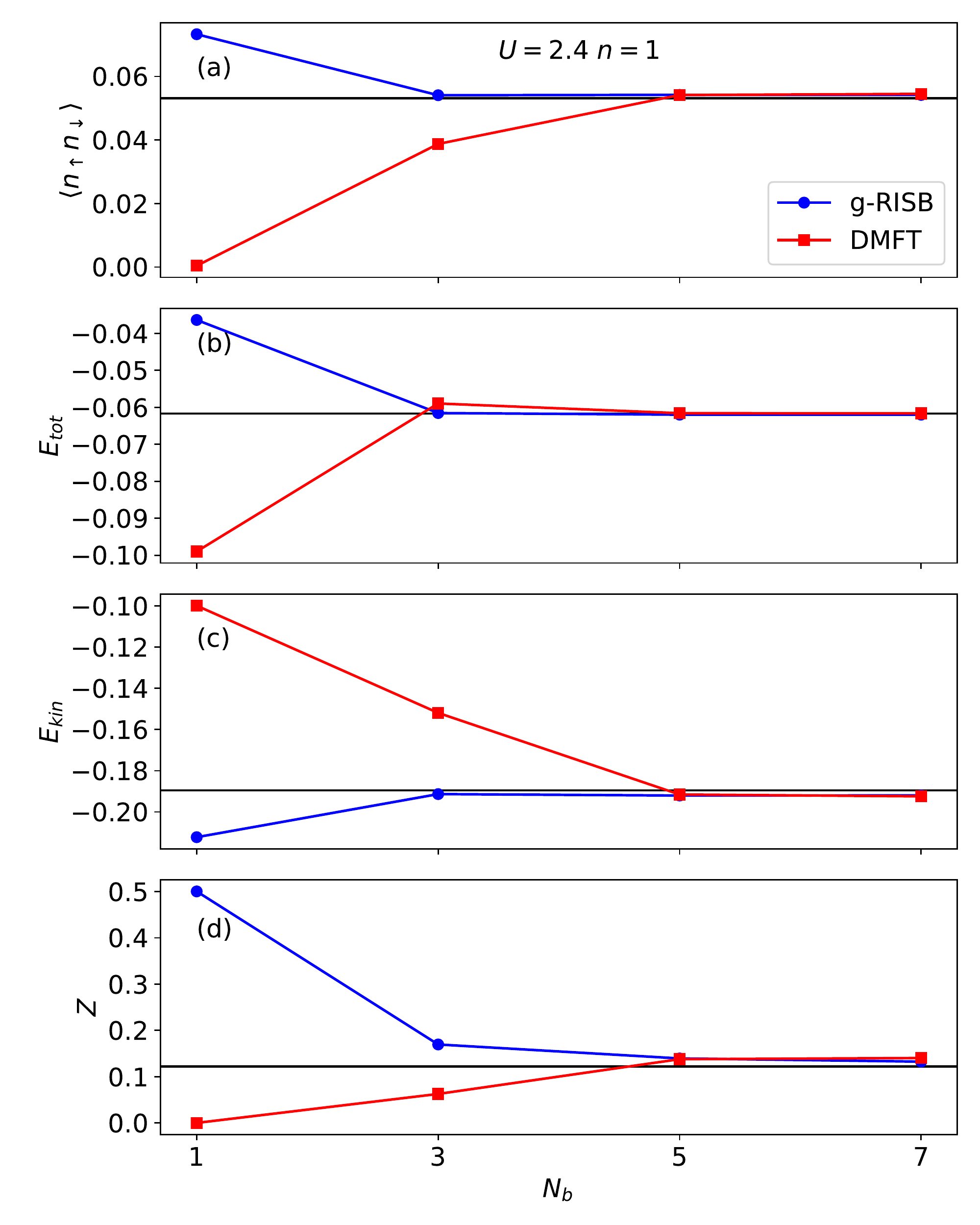}
\par\end{centering}
\caption{(a) Double-occupancy $\langle n_{\uparrow} n_{\downarrow}\rangle$, (b) total energy $E_{\text{tot}}$, (c) kinetic energy $E_{\text{kin}}$, (d) and quasiparticle weight $Z$ as a function
of bath size $N_{b}$ for $U=2.4$ at half-filling. The black horizontal
line indicates the infinite bath limit with CTQMC solver at inverse
temperature $\beta=200$. \label{fig:U2p4_half}}
\end{figure}

\begin{figure}[h]
\begin{centering}
\includegraphics[scale=0.4]{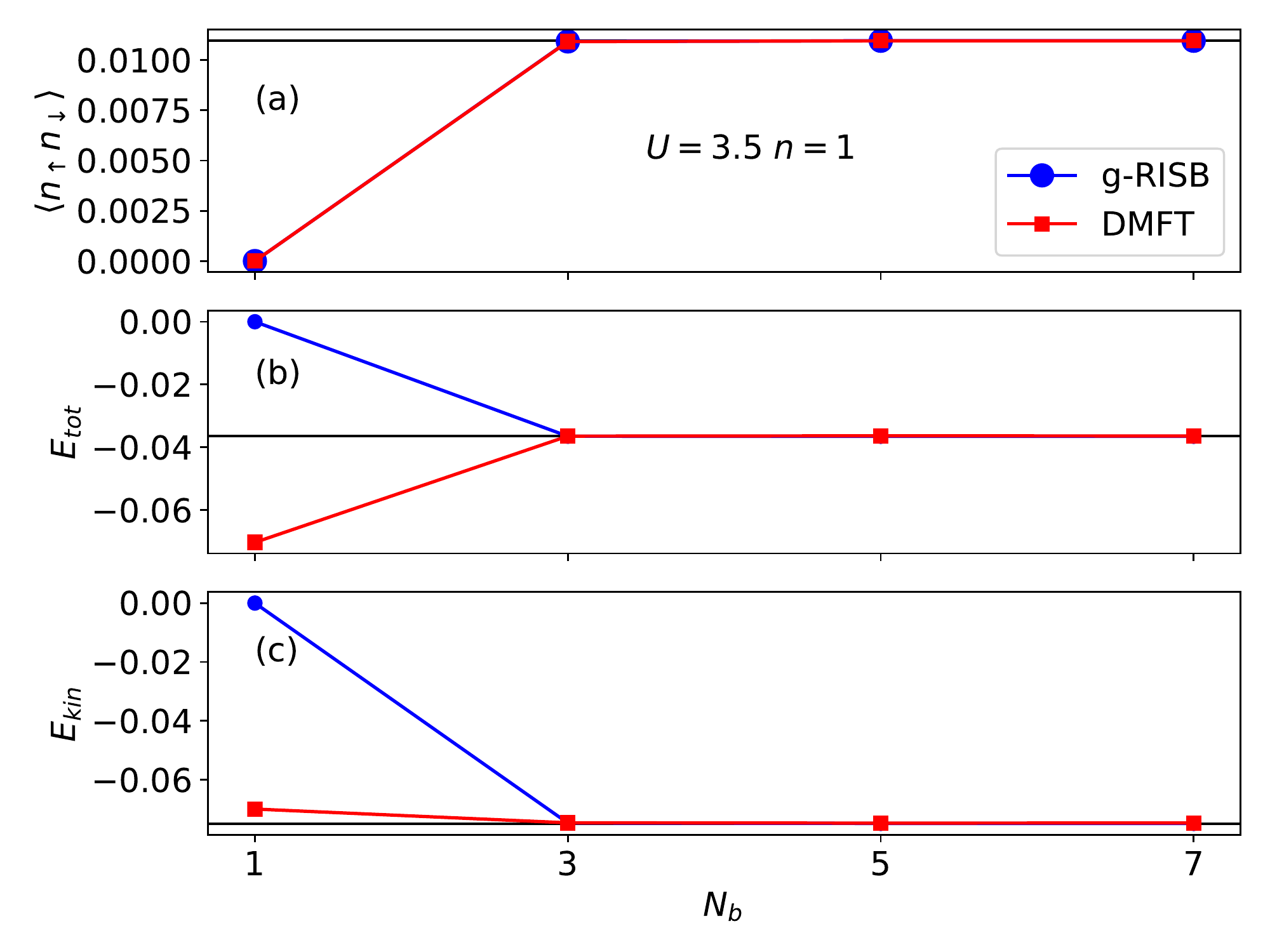}
\par\end{centering}
\caption{(a) Double occupancy $\langle n_{\uparrow} n_{\downarrow}\rangle$(b) total energy $E_{\text{tot}}$, and (c) kinetic energy
$E_{\text{kin}}$ as a function of bath size $N_{b}$
for $U=3.5$ at half-filling. The black horizontal line indicates
the infinite bath limit with CTQMC solver at inverse temperature $\beta=200$.
\label{fig:U3p5_half}}
\end{figure}

\section{Results}

\subsection{Half-filled Hubbard model}

The double occupancy $\langle n_{\uparrow}n_{\downarrow}\rangle$,
total energy $E_{\text{tot}}$, kinetic energy $E_{\text{kin}}$, and quasiparticle weight $Z$ as a function of Coulomb interaction
$U$ with the increasing numbers of bath orbitals $N_{b}$ are shown in
Fig. \ref{fig:Z_docc_E}. Our results indicate that both methods converge with less than $5\%$ error at $N_{b}=5$ for all the static physical quantities.
For all sizes of the bath, we found that g-RISB generally produces a more
accurate double occupancy and total energy closer to the converged value.

Figure \ref{fig:U2p4_half} provides us with a closer look into the convergence
behavior of the double occupancy, total energy, kinetic energy,  and the quasiparticle weight
at $U=2.4$ in the metallic phase. Our results show that the g-RISB energy and the double occupancy converge rapidly at $N_{b}=3$ with less than $1\%$ error, while
DMFT-ED requires $N_{b}=5$ to reach the same level of convergence. The quasiparticle
weight requires $N_{b}=5$ to converge to less than $5\%$ error in both methods. All the
observables converge to the DMFT solutions with the CTQMC solver at inverse temperature $\beta=200$, indicated by the black horizontal line, where the small discrepancy is originated from the finite temperature effect.

The convergence behavior in the insulating phase at $U=3.5$ is shown
in Fig. \ref{fig:U3p5_half}. We found that both methods give reliable
energy and double occupancy, and that $N_{b}=3$ yields values with errors lower than $1\%$ with respect to the CTQMC value, indicated by the black horizontal line.

\begin{figure}[t]
\begin{centering}
\includegraphics[scale=0.42]{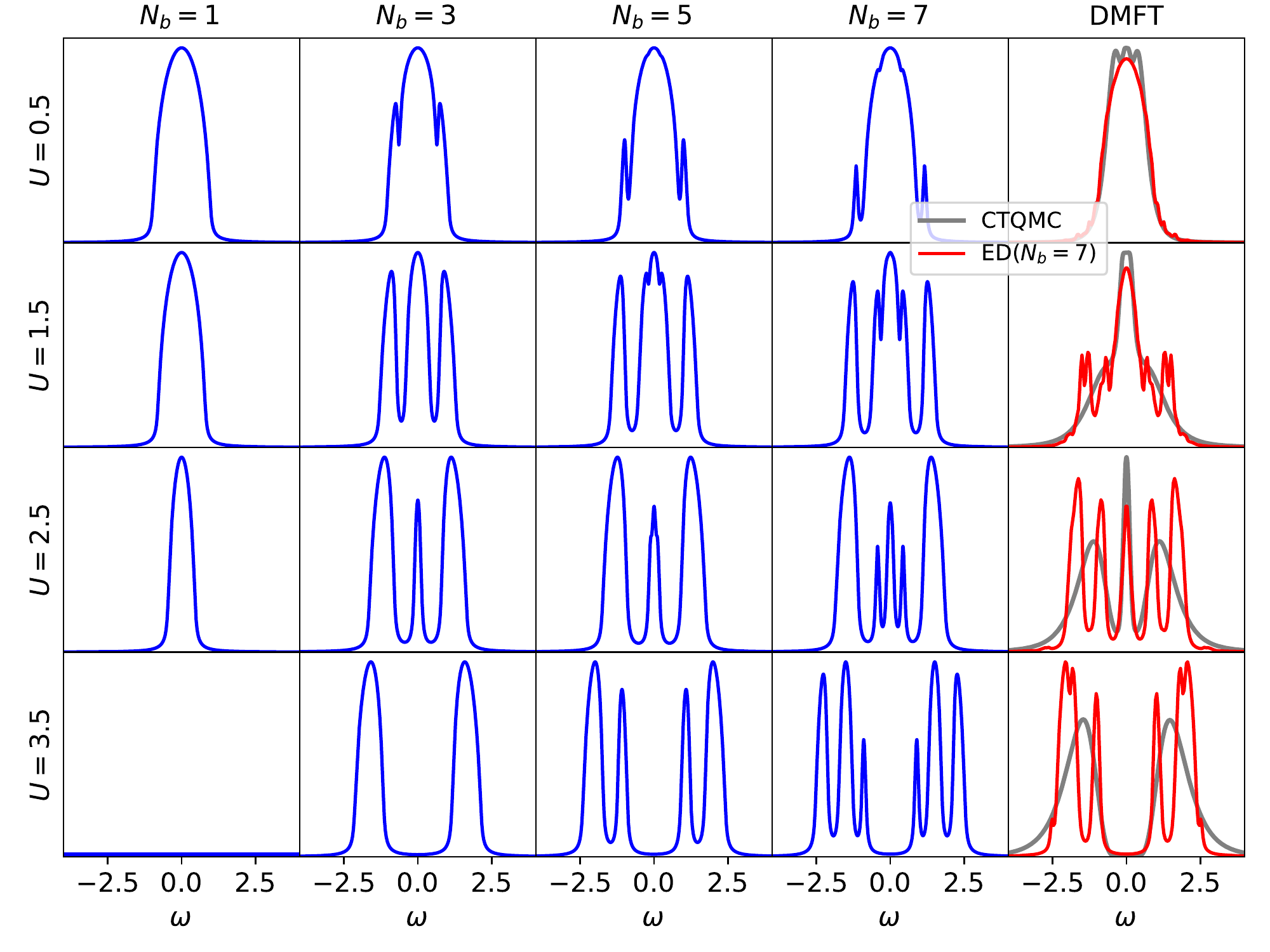}
\par\end{centering}
\caption{Density of states on Bethe lattice for different values of Coulomb interaction
$U$ and bath size $N_{b}$ (including bath and ghost-orbitals) at
half-filling. The right column is the DMFT density of states with ED (bath size
$N_{b}=7$) and the CTQMC solvers.\label{fig:DOS_half}}
\end{figure}

We now discuss the g-RISB spectral function with increasing bath size
shown in Fig. \ref{fig:DOS_half}. For $N_{b}=1$, corresponding to
the standard RISB approach, the spectral function are renormalized
following the Brinkman-Rice scenario~\cite{Brinkman1970}, where the incoherent Hubbard
bands are absent. Therefore, the density of states vanishes in the Mott
insulating phase where $Z=0$. On the other hand, with $N_{b}=3$,
g-RISB can capture reliable Mott insulator solutions where the
spectral function shows two incoherent Hubbard bands. The incoherent
Hubbard bands also emerge in the metallic spectral functions, and the
quasiparticle renormalization is significantly improved, as shown in
Fig. \ref{fig:Z_docc_E}. For $N_{b}>3$, we see the additional bath
orbitals introduce more peaks in the spectral functions. The
position of the peaks is close to the poles in the DMFT spectral
function shown in Fig. \ref{fig:DOS_half} with ED and $N_{b}=7$. The DMFT spectral functions with CTQMC solver are also shown for comparison. 

\begin{figure}[h]
\begin{centering}
\includegraphics[scale=0.4]{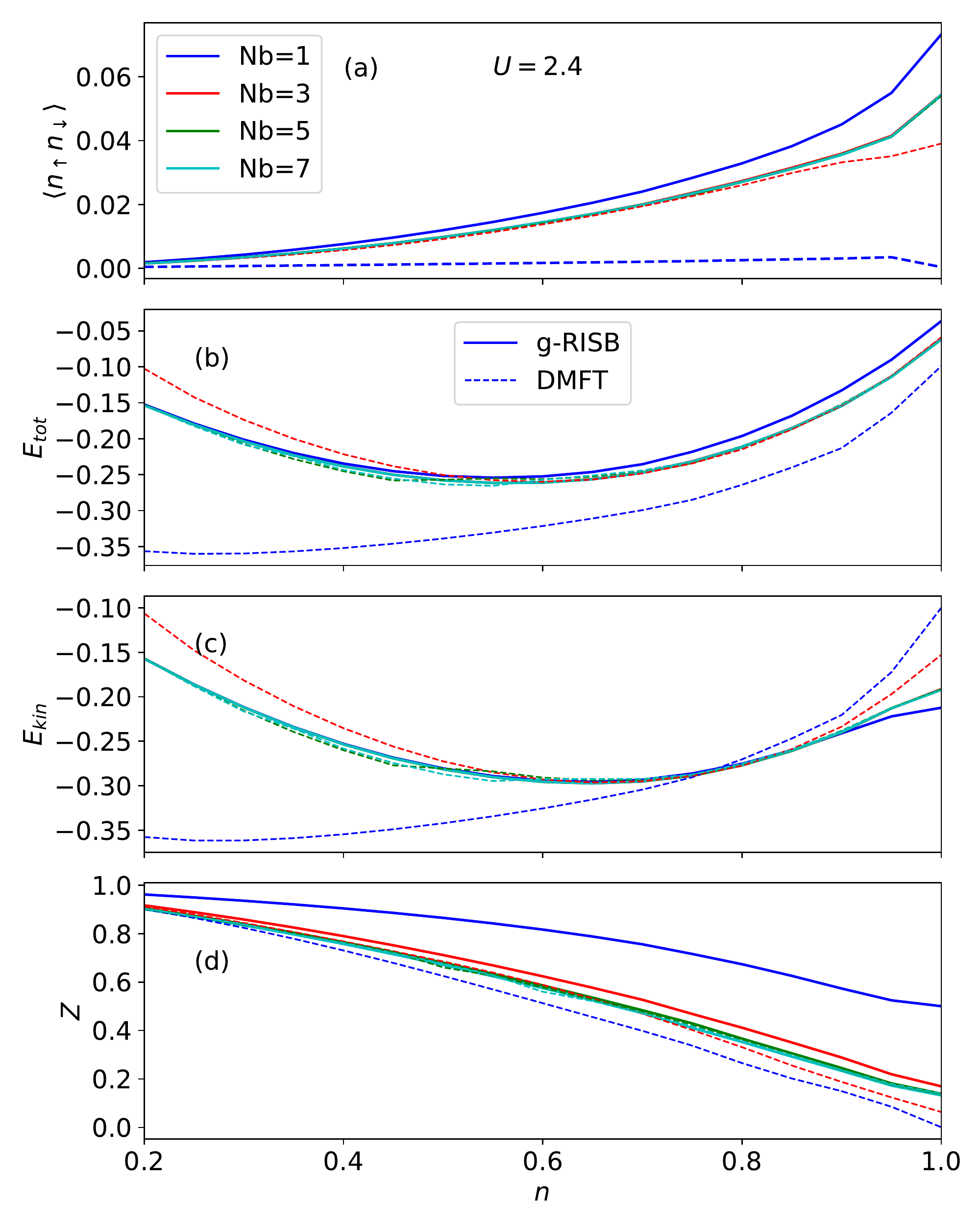}
\par\end{centering}
\caption{(a) Double occupancy $\langle n_{\uparrow}n_{\downarrow}\rangle$, (b) total energy $E_{\text{tot}}$, kinetic energy $E_{\text{kin}}$, and (d) quasiparticle weight $Z$ as
a function of electron filling $n$ with increasing bath size $N_{b}$
at $U=2.4$. The g-RISB and DMFT are shown as solid lines and dashed
lines, respectively. \label{fig:U2p4_dope}}
\end{figure}

\subsection{Doped Hubbard model}

The double occupancy, total energy, kinetic energy, and quasiparticle weight of
the doped Hubbard model as a function of filling $n$ at $U=2.4$
are shown in Fig. \ref{fig:U2p4_dope}. We found that g-RISB again produces
more accurate energies and double occupancy for all electron fillings
which are converged at $N_{b}=3$ with less than $1\%$ error. On the other hand, DMFT-ED requires $N_{b}=5$ to reach the same level of convergence. Both methods converge all physical quantities to less than $5\%$ error at $N_{b}=5$.

\begin{figure}[h]
\begin{centering}
\includegraphics[scale=0.4]{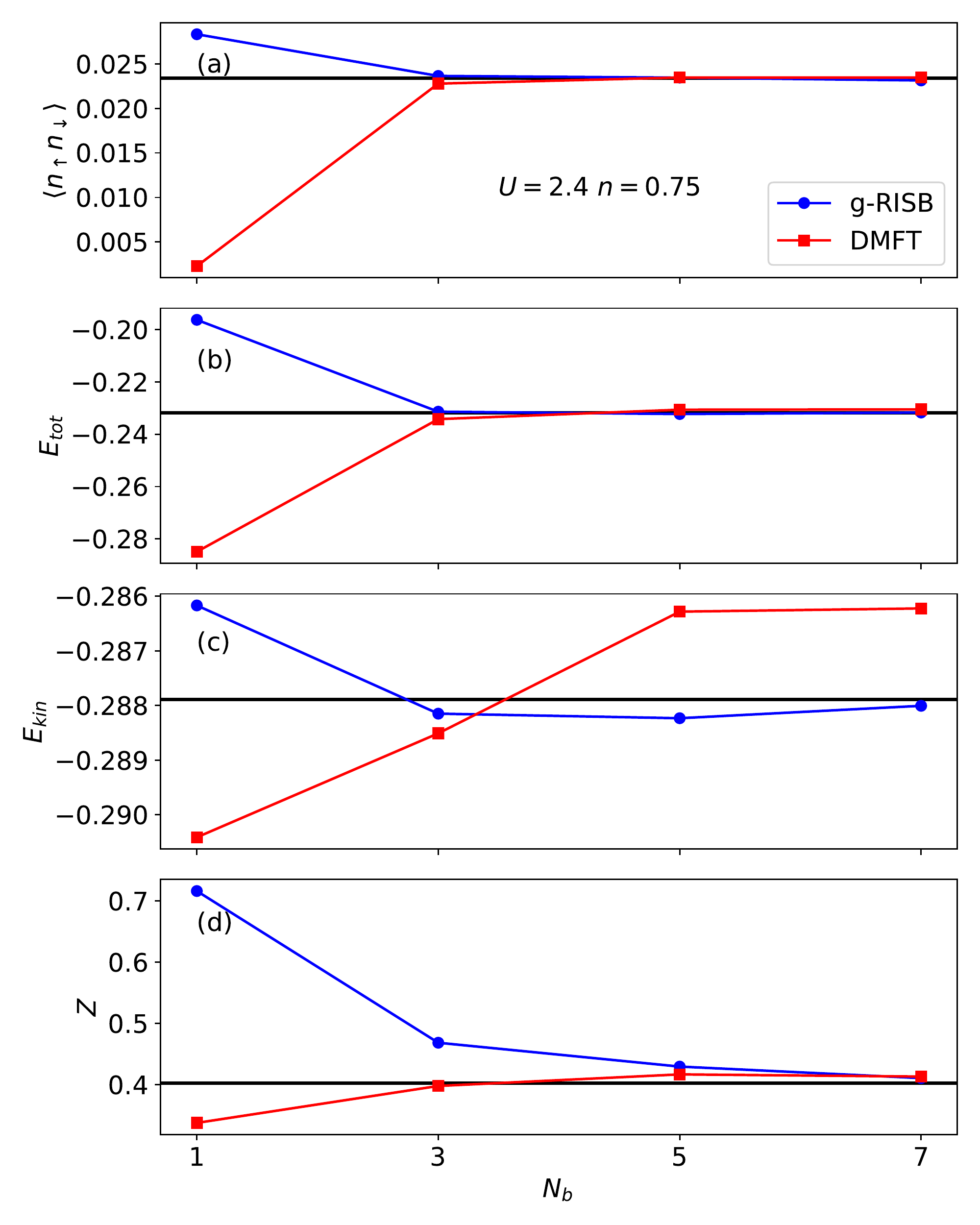}
\par\end{centering}
\caption{(a) Double-occupancy
$\langle n_{\uparrow} n_{\downarrow} \rangle$, (b) total energy $E_{\text{tot}}$, (c) kinetic energy $E_{\text{kin}}$, and (d) quasiparticle weight $Z$ as a function of bath size $N_{b}$
for $U=2.4$ at electron filling $n=0.75$. The black horizontal line
indicates the infinite bath limit with CTQMC solver at inverse temperature
$\beta=200$.
\label{fig:U2p4_n0p75_dope}}
\end{figure}

Figure \ref{fig:U2p4_n0p75_dope} shows a closer look into the convergence
behavior as a function of bath size $N_{b}$ for $U=2.4$ and filling
$n=0.75$. Our results show again that g-RISB's energy converges rapidly
with $N_{b}=3$, while DMFT-ED requires $N_{b}=5$ to reach the same accuracy.
Also, there are small differences between the DMFT-ED energy and the g-RISB energy, where the g-RISB converges to the CTQMC values indicated by the horizontal black line. The discrepancy between the DMFT-ED and the CTQMC 
results may be attributed to the effect of the artificial temperature introduced in the bath-fitting procedure. %, where the error in the energy and the chemical potential is close to the fitting error
%in the cost function. 
On the other hand, the double occupancy and
the quasiparticle weight of the two methods both converge to the CTQMC
values.

\begin{figure}[h]
\begin{centering}
\includegraphics[scale=0.42]{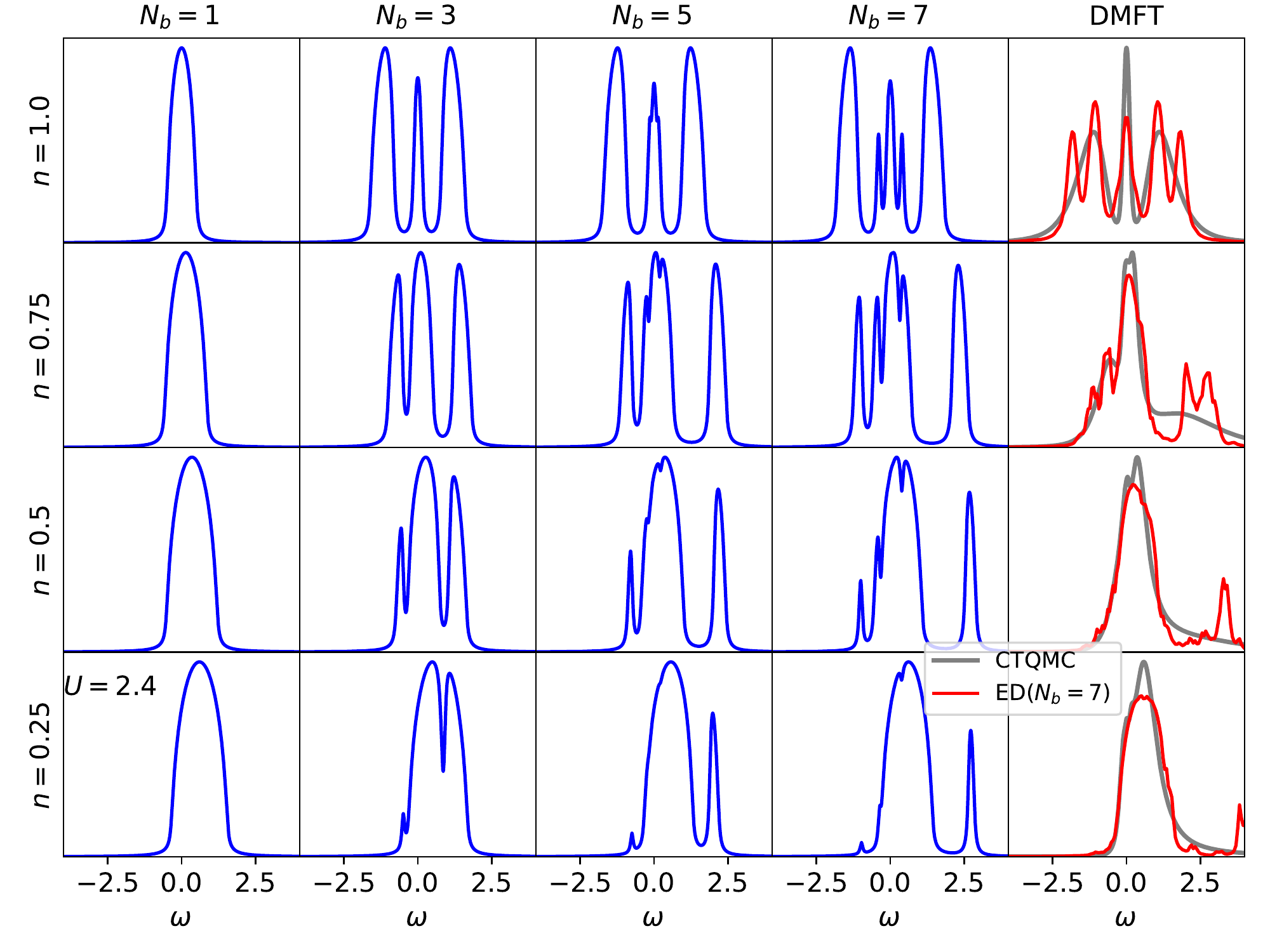}
\par\end{centering}
\caption{Density of states on Bethe lattice for different electron filling $n$
and bath size $N_{b}$ (including bath and ghost-orbitals) at $U=2.4$.
The right column is the DMFT density of states with ED (bath size $N_{b}=7$) and CTQMC solver. \label{fig:DOS_dope}}
\end{figure}

Finally, we report the spectral functions for different fillings and
bath sizes in Fig.\ref{fig:DOS_dope}. For $N_{b}=1$, we again see
that g-RISB reduces to the standard RISB approach where the incoherent
Hubbard bands are absent, and the band renormalization is of Brinkman-Rice
scenario. For $N_{b}=3$, g-RISB can capture both the quasiparticle
peak and the Hubbard bands, while the higher energy incoherent bands
are not captured. For $N_{b}>5$, we see the high energy incoherent
peaks are included and gradually shifted towards positions in agreement
with the spectral function obtained with DMFT-CTQMC (and with DMFT-ED with $N_{b}=7$ bath sites).
%spectral function with the ED solver at $N_{b}=7$ and CTQMC solver.

%\hspace{0.5cm}

\subsection{Variational property}

%It is important to note that, as explained in Refs.~\onlinecite{gRISB_2017,gRISB_2021}.
%{\color{red} 
It is important to note that the g-RISB approach~\cite{gRISB_2022} can be also formulated as a variational extension of the Gutzwiller approximation~\cite{gRISB_2017,gRISB_2021}. From this perspective, the total energy is evaluated by computing the expectation value of the Hamiltonian with respect to a variational wavefunction. Therefore, the g-RISB is variational in the limit of infinite coordination, \emph{i.e.},  the g-RISB approximation to the total energy provides us with an upper bound to the exact value. 
%}
%{\color{red} The g-RISB approach is equivalent to the multiorbital Gutzwiller approximation~\cite{gRISB_2017,gRISB_2021}, which is a variational approach in the limit of infinite coordination number~\cite{Gutzwiller1,Gutzwiller2,Gutzwiller3,Metzner_Vollhardt_1989,gRISB_2017}. Therefore, the g-RISB framework is also a variational approach in this limit, which gives the upper bound of the exact total energy that can be obtained from DMFT with the numerical exact CTQMC solver.} 
Indeed, this variational behavior is observed numerically in all of our calculations. As an example, we illustrate this in Tab.~\ref{tab:energy}) for a few cases, where we see that the total energy converges towards the exact DMFT values from above as a function of the total number of bath sites $N_b$.
%for $U=2.4$ and $n=1$ and $n=0.75$.

\begin{table}
\begin{tabular}{c c c c c c} 
%\begin{centering}
 \hline
 n\;\; & $N_b=1$\;\; & $N_b=3$\;\; & $N_b=5$\;\;  & $N_b=7$\;\; & CTQMC \\ [0.5ex] 
 \hline\hline
 1 & -0.03637 &  -0.06155 & -0.06189 & -0.06199 & -0.0621$\pm$0.0001 \\ 
 \hline
 0.75 & -0.21829 & -0.23158 & -0.23189 & -0.23190 & -0.2319$\pm$0.0001 \\
\hline
%\par\end{centering}
\end{tabular}
\caption{The g-RISB total energy at $U=2.4$ and filling $n=1$ and $n=0.75$ with different numbers of bath orbitals $N_b$. The DMFT energy at $\beta=200$ with the CTQMC solver is shown for comparison. \label{tab:energy}} 
\end{table}

\section{Conclusions}

We have compared the accuracy of g-RISB and DMFT on the single-band Hubbard model, as a function of the number of bath sites in the embedding impurity Hamiltonian. Our benchmark calculations showed that the accuracy of g-RISB can be systematically improved by increasing the number of bath sites, similar to DMFT. 
%With a few bath sites, 
Moreover, we observed that g-RISB is systematically more accurate than DMFT for the ground-state observables with a few bath sites, while the relative accuracy of these methods is generally comparable for the quasiparticle weight and the spectral function. In addition, we observed that g-RISB satisfies the variational principle in infinite dimensions, as the total energy decreases monotonically towards the exact value as a function of the number of bath sites, suggesting that the g-RISB wavefunction may approach the exact ground state in infinite dimensions.

The g-RISB only requires the ground state static density matrix for self-consistency calculations. Therefore, it circumvents the problem of evaluating the excited states (which are necessary for computing the impurity Green's function in DMFT), and opens the possibility of employing efficient ground-state wavefunction-based techniques as impurity solvers.~\cite{Zgid2012,Lin2013,Lu2014,Go2017,Mejuto-Zaera2019,Zhu2019,White1992,White1999,Chan2011,Cao_2021,Bauernfeind2017,Zhang1997,Zheng2017,Hao2019,hidden_fermion,Booth_AFQMC_EDMET}. Our results, which were obtained for a range of interaction strengths and filling factors, indicate that the g-RISB provides us with accurate energies and ground-state properties at small bath sizes, and that it has a relatively low computational cost compared to other methods.

Future research could explore the limitations and potential improvements of the g-RISB theory, and compare it with other quantum embedding approaches for more realistic systems. Overall, our results suggest that g-RISB is a promising method for first principle simulations of strongly correlated matter, which can capture the behavior of both static and dynamical observables with high accuracy, at a relatively low computational cost.

\begin{acknowledgments}
The authors thank Garry Goldstein for the careful reading of the manuscript and the useful comments. T.-H. L and G. K. were supported by the U.S. Department of Energy,
Office of Science, Office of Advanced Scientific Computing Research
and Office of Basic Energy Sciences, Scientific Discovery through
Advanced Computing (SciDAC) program under Award Number DE-SC0022198.
NL gratefully acknowledges funding from the Novo Nordisk Foundation
through the Exploratory Interdisciplinary Synergy Programme project
NNF19OC0057790.
\end{acknowledgments}

\hspace{1cm}

\appendix

%\section{Hybridization function}  \label{sec:hyb}
%
%The hybridization function is an important quantity representing the effective screening from the lattice. Figure. ~\ref{fig:hyb} shows the hybridization functions $\Delta(\omega)$ computed from g-RISB and DMFT with ED and CTQMC solvers. The g-RISB hybridization function is computed from $\Delta_\sigma(\omega)=\sum_{a} |D_{a\sigma}|^2/(\omega+i0^+ +\lambda^c_{aa})$, where we performed gauge transformation so that $\lambda^c$ is diagonal~\cite{Lecherman_2007,Lanata_2017_PRL}. We found good agreement in the location of the poles between g-RISB and DMFT-ED calculations, which discretizes the DMFT-CTQMC continuous hybridization function.
%
%\begin{figure}[H]
%\begin{centering}
%\includegraphics[scale=0.42]{hybw}
%\par\end{centering}
%\caption{ Imaginary part of the hybridization function $-\text{Im}\Delta(\omega)$ at (a) $U=2.4$ and $n=1$ and (b) $U=2.4$ and $n=0.75$ for g-RISB, DMFT-ED and DMFT-CTQMC. \label{fig:hyb}}
%\end{figure}

\bibliographystyle{apsrev}
\bibliography{ref}

\end{document}